\documentclass[apj]{emulateapj}
\graphicspath{{figures/}}
\DeclareGraphicsExtensions{.jpg,.pdf,.png,.eps,.ps}

\usepackage[table,usenames,dvipsnames]{xcolor}
\usepackage[backref,breaklinks,colorlinks,citecolor=blue]{hyperref}
\usepackage{natbib}
\usepackage{soul}

\newcommand{\cpmftens}{\ensuremath{C_{\ell}^{\rm PMF,\,tens}}}
\newcommand{\cpmfvec}{\ensuremath{C_{\ell}^{\rm PMF,\,vec}}}
\newcommand{\apmf}{\ensuremath{A_{\rm PMF}}}
\newcommand{\bpmf}{\ensuremath{B_{\rm 1\,Mpc}}}
\newcommand{\alens}{\ensuremath{A_{\rm lens}}}
\newcommand{\lcdm}{\ensuremath{\Lambda}CDM}
\newcommand{\nrun}{\ensuremath{n_{\rm run}}}
\newcommand{\neff}{\ensuremath{N_{\rm eff}}}
\newcommand{\ho}{H\ensuremath{_0}}
\newcommand{\mnu}{\ensuremath{\sum m_\nu}}
\newcommand{\ukarcmin}{\ensuremath{\mu}{\rm K-arcmin}}

\newcommand{\fermilat}{\textit{Fermi}-LAT}

\newcommand{\be}{\begin{equation}}
\newcommand{\ee}{\end{equation}}
\newcommand{\planck}{{\sl Planck}}

\newcommand{\bicepkeck}{BICEP2/Keck Array}
\newcommand{\sptnew}{SPT-3G}
\newcommand{\pb}{\textsc{Polarbear}}
\newcommand{\simons}{Simons Array}
\newcommand{\sptpol}{SPTpol}
\newcommand{\advactpol}{Adv.~ACTpol}

\newcommand{\changed}[1]{\textcolor{Black}{#1}}
\newcommand{\removed}[1]{\textcolor{Black}{}}

\newcommand{\upperAall}{ 0.36}
\newcommand{\upperAplk}{ 0.76}
\newcommand{\upperAplkalens}{ 0.64}
\newcommand{\upperAallalens}{ 0.30}
\newcommand{\upperAsptpol}{ 0.45}
\newcommand{\upperAbicep}{ 0.44}

\newcommand{\upperAallalensr}{ 0.25}
\newcommand{\medAlensall}{ 1.123}
\newcommand{\sigmaAlensall}{ 0.064}
\newcommand{\medAlensRall}{ 1.127}
\newcommand{\sigmaAlensRall}{ 0.063}
\newcommand{\upperRall}{ 0.07}

\newcommand{\upperBplk}{ 1.8}
\newcommand{\upperBall}{ 0.91}

\newcommand{\fisherAplk}{0.38} 
\newcommand{\fisherASThree}{0.020}

\newcommand{\fisherAlooseNbOneplk}{0.38} 
\newcommand{\fisherAlooseNbTwoplk}{\ensuremath{3.0\times10^{-6}}} 
\newcommand{\fisherAlooseSThree}{0.022} 
\newcommand{\fisherAlooseNbOneSThree}{0.023} 
\newcommand{\fisherAlooseNbTwoSThree}{\ensuremath{1.5\times10^{-7}}} 

\newcommand{\fisherAlooseSFour}{\ensuremath{6.3\times10^{-3}}} 
\newcommand{\fisherAlooseNbOneSFour}{\ensuremath{7.4\times10^{-3}}} 
\newcommand{\fisherAlooseNbTwoSFour}{\ensuremath{5.2\times10^{-8}}} 

\def\Melbourne{1}
\def\uci{2}
\begin{document}

\title{Current and future constraints on primordial magnetic fields}
\author{Dylan~R.~Sutton\altaffilmark{\Melbourne}, Chang~Feng\altaffilmark{\uci}, and Christian~L.~Reichardt\altaffilmark{\Melbourne}}
\altaffiltext{\Melbourne}{School of Physics, University of Melbourne, Parkville, VIC 3010, Australia}
\altaffiltext{\uci}{Department of Physics and Astronomy, University of California, Irvine, CA 92697-4575, USA}
\email{christian.reichardt@unimelb.edu.au}

\begin{abstract}

We present new limits on the amplitude of potential primordial magnetic fields (PMFs) using temperature and polarization measurements of the cosmic microwave background (CMB)  from \planck{}, \bicepkeck{}, \pb, and \sptpol. 
We reduce twofold the 95\% CL upper limit on the CMB anisotropy power due to a nearly-scale-invariant PMF, with an allowed B-mode power at $\ell=1500$ of $D_{\ell=1500}^{BB} < 0.071 \mu K^2$ for \planck{} versus $D_{\ell=1500}^{BB} < 0.034 \mu K^2$ for the combined dataset. 
We also forecast the expected  limits from soon-to-deploy CMB experiments (like \sptnew{},  \advactpol, or the \simons) and the proposed CMB-S4 experiment. 
Future CMB experiments should dramatically reduce the current uncertainties, by one order of magnitude for the near-term experiments and two orders of magnitude for the CMB-S4 experiment. 
The constraints from CMB-S4 have the potential to rule out much of the parameter space for PMFs.
\end{abstract}

\section{Introduction}
\label{sec:intro}

Measurements of the cosmic microwave background (CMB) temperature anisotropy have provided some of the most powerful tests of cosmology. 
We are now entering a new era as experiments begin to measure  polarized ``B-modes" in the CMB for the first time \citep{hanson13,polarbear14b,bicep14a,naess14,keisler15,bicepkeck15,louis16}. 
Precision measurements of CMB polarization promise new tests of the standard cosmological model. 
The best known of these tests are the searches for inflationary gravitational waves in B-modes at large angular scales and plans to measure the sum of the neutrino masses through the lensing B-modes on small angular scales \citep[for a recent review see,][]{abazajian16}. 

CMB B-mode measurements can also be used to constrain more exotic models, such as the possible existence of cosmic birefringence \citep{carroll98,lue99} or primordial magnetic fields (PMFs) \citep{kosowsky96, seshadri01}.  
Both effects lead to the rotation of E-modes into B-modes, thereby increasing the amplitude of the BB power spectrum.
By equating the magnitude of the resulting B-modes, parity-violating processes can be translated into an equivalent PMF strength. 
It is therefore common to simply quote effective limits on PMFs. 
In this work, we present new upper limits from current CMB polarization data on the possibility of PMFs or parity-violating physics. 

Magnetic fields are ubiquitous in astronomy and are found almost universally in collapsed objects from stars to galaxies and galaxy clusters \citep[for a review, see][]{ryu12, widrow12}. 
There is even some evidence for magnetic fields in intergalactic space from \fermilat{} data \citep{neronov10}.
High energy $\gamma$-rays from blazars should  produce electron-positron pairs when the $\gamma$-rays collide with IR or optical photons. 
These pairs should later annihilate at GeV energies, but the expected GeV flux is missing in the \fermilat{} observations. 
The missing flux could be explained by the deflection of particles due to intergalactic magnetic fields. 
There are, however, alternative explanations involving instabilities in the electron-positron plasma, leading pairs to deposit their energy in the intergalactic medium in the form of heat \citep{broderick12}.
If the loss of flux is indeed due to magnetic fields, then the GeV results set lower limits on the intergalactic magnetic field strength of $10^{-9} - 10^{-6}$\,nG \citep{tavecchio10,taylor11,dermer11,vovk12}.

The mechanism to create large-scale magnetic fields, especially in intergalactic space, remains unclear. 
One popular proposal is that the observed fields are the product of PMFs, which are predicted by several theories of the early Universe \citep[e.g.,][]{turner88, grasso98,ichiki06}. 
Adiabatic compression and turbulent shocks during later structure formation amplify these initial seed PMFs into the stronger magnetic fields observed today. 
Of course, this amplification process may be seeded through other means, such as AGN or galactic dynamos \citep[for a review, see][]{giovannini04}. 
However, the possibility of PMFs opens up the intriguing idea that observations of large-scale magnetic fields may offer insights into the physics of the very early Universe.

PMFs would have observational consequences for Big Bang nucleosynthesis \citep[e.g.,][]{kahniashvili10}, large scale structure \citep[e.g.,][]{battaner97}, the ionization history of the Universe \citep{kunze15}, and the black body spectrum of the CMB \citep[e.g.,][]{kunze14},  as well as the CMB anisotropies. 
The CMB anisotropies have yielded some of the strongest constraints on PMFs and are the focus of this work.
There have been three recent results of note. 
\citet{planck15-19} have used the \planck{} 2015 release of temperature and polarization data to set limits on a variety of PMF models. 
With the CMB power spectrum data that will be the focus of this paper, \citet{planck15-19} find 95\% CL upper limits ranging from $\bpmf < 5.6$\,nG to $<0.7$\,nG depending on the exact model. 
Here \bpmf{} is the rms magnetic field strength on a 1\,Mpc length scale. 
The \pb{} collaboration also recently announced limits on PMFs from the \pb{} data using either a 4-point estimator or B-mode power spectrum measurement \citep{polarbear15}. 
The strongest constraints were from the B-mode spectrum; the observed upper limit was $\bpmf < 3.9$\,nG. 
Two other experiments, \sptpol{} and \bicepkeck{}, have also reported B-mode power spectrum measurements \citep{keisler15,bicepkeck15}, 
and \citet{zucca16} reported constraints from the combination of the \planck{}  and \sptpol{} bandpowers. 
In this paper, we will combine the data from all four experiments to determine upper limits on PMFs, and find that the \bicepkeck{} data in particular improves these upper limits. 
We then present Fisher matrix forecasts on PMF models for the stage-III and stage-IV CMB experiments being built or designed right now.

The outline of this paper is as follows. 
In \S\ref{sec:data}, we lay out the data and how we use the data to put limits on PMFs. 
We present the results of this analysis on current data in \S\ref{sec:results}. 
In \S\ref{sec:forecasts}, we forecast constraints from future experiments using a Fisher matrix formulism. 
We conclude in \S\ref{sec:conclusions}. 

\section{Data and Methods}
\label{sec:data}

We use  Markov chain Monte Carlo (MCMC) methods to study constraints on PMFs. 
In this section, we first describe the CMB temperature and polarization anisotropy data used, and then discuss the MCMC implementation.

\subsection{Data}

\begin{figure*}[htb]\centering
\includegraphics[width=0.9\textwidth,clip,trim={1.5cm 12.5cm 5cm 3.8cm}]{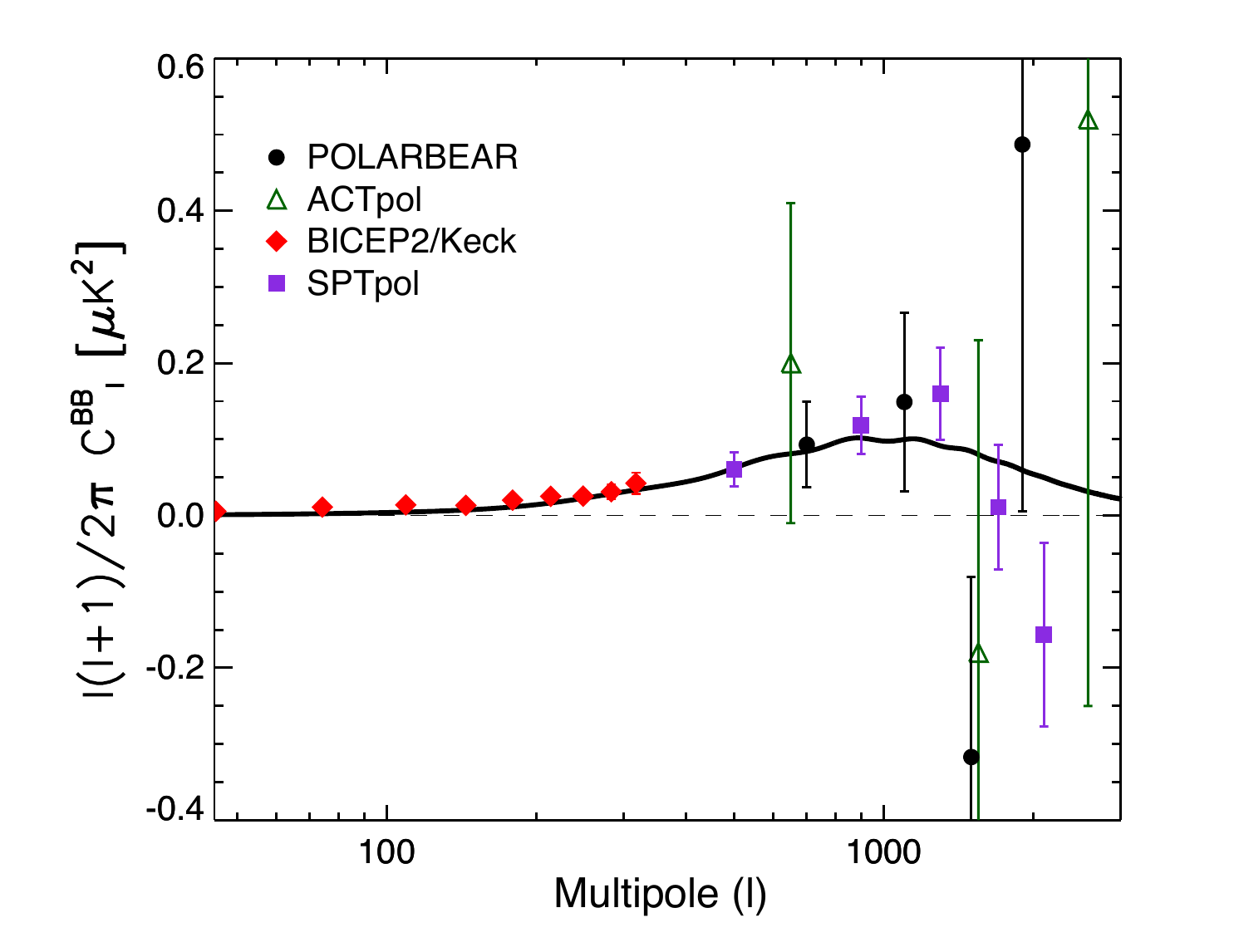}
  \caption[Current measurements of CMB B-modes]{
  Current measurements of the CMB B-mode power spectrum. 
    The black line shows the expected B-mode power spectrum in the \planck{} best-fit \lcdm{} cosmology (with the tensor-to-scalar ratio, r set to 0.01). 
  The data sets are from POLARBEAR (black circles; \citet{polarbear14b}); ACTpol (dark green triangles; \citet{naess14});  BICEP2/Keck Array (red diamonds; \citet{bicepkeck15}); and SPTpol (purple squares; \citet{keisler15}). 
      The B-mode measurements used for the limits in this work are denoted with filled symbols. 
      Current B-mode measurements are noise-limited at all angular multipoles; the next generation of experiments will significantly increase the signal-to-noise on the B-mode power spectrum. 
           \label{fig:pmf-experiments}
  }
\end{figure*}

We use a compendium of current measurements of the CMB temperature and polarization anisotropies from ground-based and satellite experiments. 
We use \planck{} data from the 2015 release to constrain the TT, TE, EE and lensing power spectra. 
Specifically, these are the ``plik\_dx11dr21\_HM\_v18\_TT", ``lowTEB" and ``lensing" \planck{} likelihood modules.

In addition to the \planck{} data, we use a number of ground-based CMB B-mode polarization measurements. 
First, we include measurements of the TE, EE and BB power spectra from the \sptpol{} experiment \citep{crites15,keisler15}. 
The BB bandpowers cover the angular multipoles $\ell \in [500,2000]$. 
We also add the BB bandpowers from \pb{} that cover the multipoles from 500 to 2500 \citep{polarbear15}. 
At these angular scales, both the \sptpol{} and \pb{} bandpowers primarily constrain the vector modes of a PMF. 
Finally, we include the latest BICEP2 and Keck Array  joint analysis \citep{bicepkeck15}. 
This last dataset also places limits on the tensor modes of a PMF due to its coverage of lower multipoles. 
During the writing of this work, ACTpol polarization power spectra became available \citep{naess14,louis16}. 
We have not included them as the public likelihoods do not include the B-mode power spectrum; we do not expect adding the ACTpol bandpowers to significantly change the PMF limits based on a visual comparison of the current BB bandpowers from different experiments (see Fig.~\ref{fig:pmf-experiments}).
In all chains, we marginalize over the recommended foreground models for each data set. 
We do not however require consistency between these foreground models as the data don't have identical flux cuts for extragalactic sources and so on. 
These data are plotted against the expected B-mode power spectrum for the standard cosmological model in Fig.~\ref{fig:pmf-experiments}. 

\subsection{Methods}

We use MCMC methods to determine the parameter constraints reported in this work. 
The results are calculated using  the {\textsc CosmoMC}\footnote{http://cosmologist.info/cosmomc (July 2015 version)} package \citep{lewis02b}. 
CosmoMC invokes  CAMB\footnote{http://camb.info}  \citep{lewis00} to calculate the CMB power spectrum for each set of cosmological parameters. 
Although CosmoMC and CAMB have a partial implementation of PMFs, we choose to adopt a simpler, fast, template-based calculation for the PMF-sourced power. 
We have adapted CosmoMC to add a scaled version of a PMF template to all four CMB power spectra: TT, TE, EE, and BB, where the scale factor is \apmf. 
\be
\changed{C_\ell = C_\ell^{\rm{CAMB}} + \apmf \left[\cpmfvec + \left(\frac{\beta}{20}\right)^{1.9}  \cpmftens \right]}\label{clmodel}
\ee
\changed{The calculation of the PMF templates, \cpmfvec{} and \cpmftens{}, and the motivation for the $\beta$ scaling are described in \S\ref{sec:template}. 
The parameter $\beta = {\rm ln}(a_{\nu}/a_{\rm{PMF}})$ relates the timing of neutrino decoupling and the generation of the PMF, where $a_x$ represents the scale factor at the respective events. }
Note that no other effect of  PMFs is considered; this is one reason we only use CMB data.

For the real data, unless noted, we assume the six-parameter, spatially-flat \lcdm{} model with a single massive neutrino of 60\,meV and \changed{2 parameters, \apmf{} and $\beta$, }describing the power due to PMFs. 
We adopt flat priors on all parameters. 
\changed{There are two points to note about the PMF priors. 
First, this is a flat prior on the observed PMF power, $\apmf$, not the rms magnetic field strength, \bpmf, that has often been used in the literature. 
In the current era of upper limits, this prior choice has some impact on the resulting PMF limits. 
Second, the exact range of the uniform prior on $\beta$ matters. 
Here we follow \citet{planck15-19} and \citet{zucca16} and use $\beta \in [11.513, 41.447]$.\footnote{This corresponds to $log_{10} (a_{\nu}/a_{\rm{PMF}}) \in[4,17]$.}
However, the constraints, especially those from \planck{} alone, weaken substantially if the lower bound on $\beta$ is lowered further. 
If we follow \citet{polarbear15} who used $\beta \in [0,39]$, the upper limits are relaxed by a factor of roughly eight for \planck-only and by a factor of two for the combined dataset. }
We sometimes also add \alens{} as a simple way of marginalizing over uncertainty in the predicted lensed BB power spectrum for any extension to the \lcdm{} model. 
Finally, we sometimes allow non-zero tensors, parameterized by the tensor-to-scalar ratio, r. 

\subsection{Primordial Magnetic Field Template}
\label{sec:template}

\begin{figure*}[htb]\centering
\includegraphics[width=0.9\textwidth,clip,trim={.5cm 12.cm 6cm 5cm}]{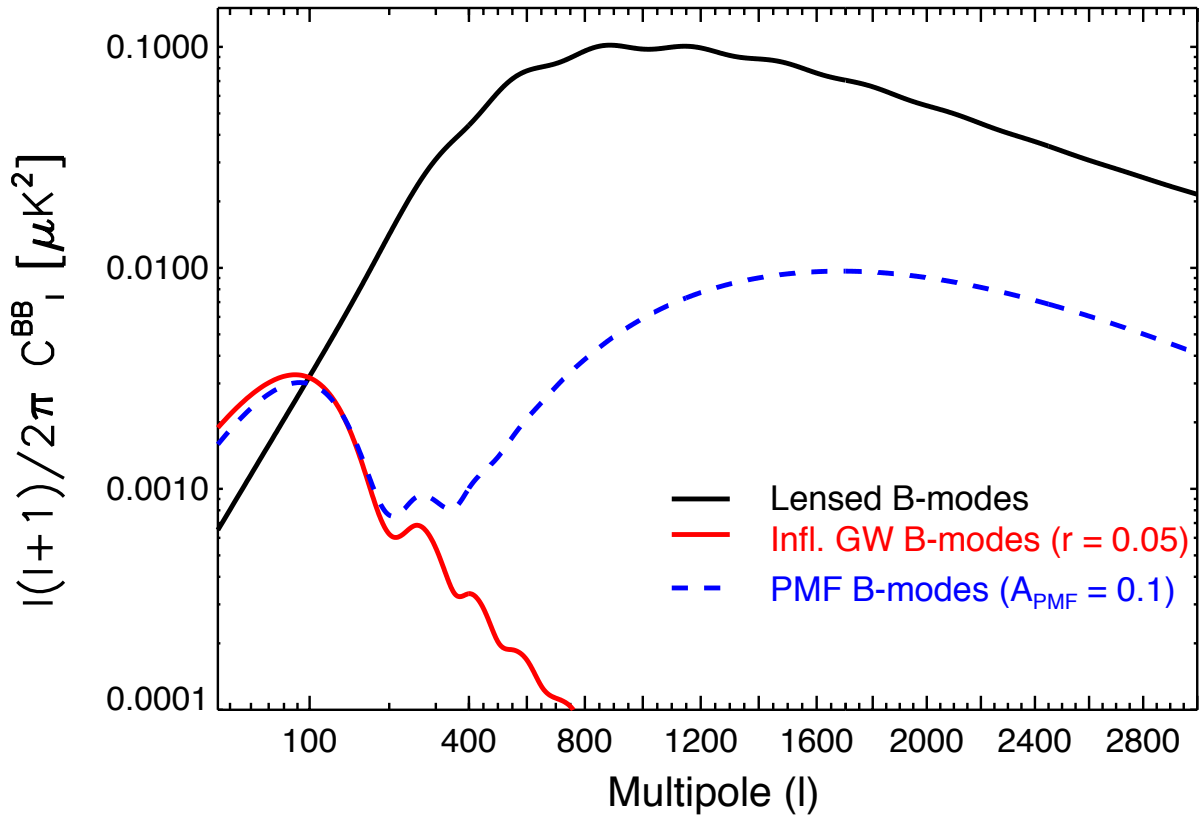}
  \caption[CMB polarization from PMFs]{
  The expected sources of CMB B-modes.
  The black line shows the expected lensing B-mode power in the \planck{} best-fit \lcdm{} cosmology, while the red line corresponds to the B-modes due to inflationary gravitational waves for $r=0.05$. 
  The dashed blue line shows the nearly-scale invariant ($n_B=-2.9$) PMF template used in this work, scaled to \apmf{} = 0.2 (i.e.~\bpmf{}=1.7\,nG) to match the gravitational wave signal. 
  At $\ell < 200$, the tensor contribution from PMFs closely resembles the expected inflationary gravitational wave signal, and the two signals are effectively degenerate on these angular scales in a power spectrum (2-point) analysis.
      \label{fig:pmf-bb}
  }
\end{figure*}

The density and stress perturbations introduced by PMFs give rise to vorticity and gravitational waves in the early photon-baryon  plasma, i.e., vector and tensor modes, as well as scalar modes. 
The exact amplitude of these modes can be calculated by modifying the normal Boltzmann equations to include sources due to the PMFs. 
The primary PMF anisotropy is assumed to be Gaussian distributed with a power law power spectrum, $A_B k^{n_B}$, where $n_B$ is the spectral index and $A_B$ is the PMF power normalization. 
We can translate the normalization, $A_B$, into more physically-motivated units by calculating the rms of the PMF field strength over a length scale of 1\,Mpc, which we will call \bpmf. 
\changed{
Finally, the timing of when PMFs are generated relative to neutrino decoupling matters for the tensor component because the PMF-induced stress anisotropy can be compensated by decoupled or partially-decoupled neutrinos. 
The timing is  parametrized by $\beta = ln(a_{\nu}/a_{\rm{PMF}})$,  the natural logarithm of the ratio of the scale factors at neutrino decoupling and PMF generation. 
}

\begin{figure}[htb]\centering
\includegraphics[width=0.45\textwidth,clip,trim={1.5cm 12.cm 5.5cm 4cm}]{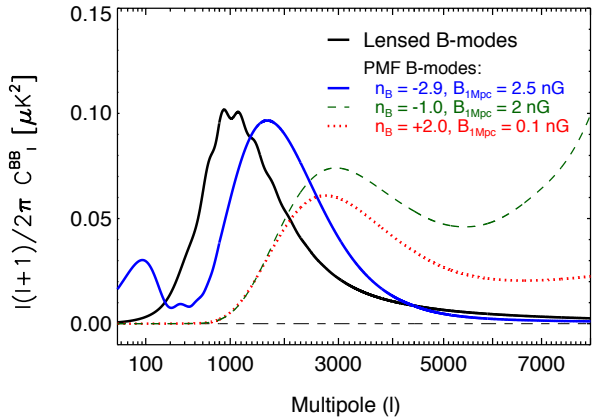}
  \caption[CMB polarization from PMFs with different spectral indices]{ \label{fig:pmf-nb}
  How the expected B-mode power due to PMFs changes with the spectral index, $n_B$. 
  The lensing B-mode power spectrum (solid, black line) is plotted against the PMF power for $n_B=-2.9$ (solid, blue line), $n_B=-1$ (dashed, green line), and $n_B=+2$ (dotted, red line). 
  To keep all the curves on the same scale, we adjust the RMS strength of the magnetic field, \bpmf{}, between the three cases as noted in the legend. 
  For a fixed RMS magnetic field strength, increasing $n_B$ greatly increases the expected B-mode power. 
  There is an inflection point between the nearly-scale-invariant PMF with $n_B=-2.9$ and the bluer PMF spectra ($n_B=-1$ or 2) which shifts power to smaller angular scales and eliminates the peak at $\ell \lesssim 200$. 
  }
\end{figure}

Given that we are still in the era of upper limits, we \changed{fix the value of $n_B$ instead of exploring the full 3-dimensional PMF parameter space. 
For the real data, we focus on nearly-scale-invariant ($n_B = -2.9$) PMF template in this work due to its connection to inflation.  
Quantum mechanical zero-point fluctuations during inflation can generate an approximately scale-invariant PMF with a spectral index close to $n_B = -3$ \citep[e.g.,][]{turner88, mack01,2008PhRvD..78f3012K}. 
In contrast, causal PMFs with bluer spectra require later phase transitions or other new physics after inflation \citep[e.g.,][]{durrer03, subramanian16}. 
We also consider $n_B\,=\,-1$ and $n_B\,=\,2$ for the forecasts in \S\ref{sec:forecasts}. 
For each value of $n_B$, we use publicly released modifications to CAMB from \citet{zucca16} to calculate a PMF template for the CMB TT, TE, EE and BB power spectra at fixed parameters: \bpmf = 2.5\,nG and $\beta=20.72$ ($a_{\nu}/a_{\rm{PMF}} = 10^9$).} 
The calculated templates include tensor and vector PMF contributions; scalar contributions are not implemented. 
The lack of scalar terms should not matter for CMB polarization data. 
Temperature data is not the main focus of this work, and we find our \planck{} results  to be consistent with \citet{planck15-19}, suggesting the impact of the scalar terms is minor. 
The PMF templates are plotted against other signals in the CMB B-mode power spectrum in Fig.~\ref{fig:pmf-bb} \changed{and against each other in Fig.~\ref{fig:pmf-nb}. }

As noted in the last section, the MCMC includes a  scaling parameter, \apmf{}, for the PMF power spectrum template. 
We can translate the amplitude constraint on this template to a constraint on the r.m.s. magnetic field strength on 1\,Mpc scales, \bpmf, using the expected scaling:
\be \label{eqn:scaling}
C_{\ell}^{\rm{vec}} \propto \bpmf^4.
\ee
An unfortunate consequence of this steep fourth-power scaling is that substantial reductions in the observationally allowed power lead to only modestly better limits on the magnetic field strength. 
However, if a PMF is detected, the steep scaling between \bpmf{} and the observed CMB B-mode power spectrum would allow excellent constraints on the PMF properties.

\changed{The power in the tensor modes also depends on the timing of PMF generation relative to neutrino decoupling, and is expected to scale approximately as $\beta^2$ \citep{lewis04,shaw10b}.   
We double-check this expectation for $\beta$ in the range 2 to 20 using the  \citet{zucca16} code, and find the computed tensor power scales as $\beta^{1.9}$
We use this $\beta$ power law form to handle the PMF timing dependence. 
Thus the tensor PMF power will scale as:
\be \label{eqn:betascaling}
\apmf^{tens} \propto  \bpmf^4 \beta^{1.9},
\ee
}

\section{Results}
\label{sec:results}

We begin by investigating how the addition of ground-based CMB polarization data to \planck{} data affects the limits on PMFs. 
For the real data, we focus on the most theoretically motivated scenario of an approximately scale-invariant spectrum for the PMFs ($n_B=-2.9$), although we will consider other scenarios in the forecasts of the next section. 
The ground-based CMB polarization data  substantially reduce the allowed PMF power; we see a factor of two reduction over \planck{} alone in the simple \changed{8}-parameter model. 
The 95\% confidence upper limit for \planck{} is $\apmf <  \upperAplk$ for \lcdm{}+PMF. 
Adding the ground-based experiments brings this limit to $\apmf <  \upperAall$. 
To facilitate a physical interpretation, we can restate these limits as an allowed B-mode power  at $\ell=1500$, which is near the peak of the PMF contribution. 
The equivalent limits on the allowed B-mode power are $D_{\ell=1500}^{BB} < 0.071 \mu K^2$ for \planck{}-alone and $D_{\ell=1500}^{BB} < 0.034 \mu K^2$ when the ground-based experiments are added. 
We test which datasets are important by adding single datasets to the \planck{} data, \changed{and find that the source of the improvement is evenly split between \sptpol{} and \bicepkeck{}. 
Adding either one to \planck{} yields the same 95\% CL upper limit: $\apmf^{\planck+\bicepkeck} <  \upperAbicep$ versus  $\apmf^{\planck+\sptpol}) < \upperAsptpol$. 
Extending the $\beta$ prior to lower values would shift the relative weight more towards \sptpol{} (and vice-versa). 
Similarly, increasing the value of $n_B$ would  increase the relative importance of the smaller angular scales measured by \sptpol.
In short, the source of the factor of two reduction in the upper limit on \apmf{} for a nearly-scale-invariant PMF spectrum ($n_B=-2.9$) is due to a more accurate measurement of the B-mode power spectrum at all scales. }

We next look at the model dependence of these upper limits. 
We list the limits for the combined dataset  with different parameter sets in Table~\ref{tab:param_all}. 
First, we vary the largest source of  B-mode power, gravitational lensing, by allowing \alens{} to float. 
Somewhat counterintuitively, the limits are \changed{somewhat} better with \alens{} allowed to vary (this is due to a mild preference for \alens{} above unity which increases the lensed B-mode power). 
In this case for  Planck alone, the 95\% confidence upper limit is $\apmf <  \upperAplkalens$, dropping to $\apmf <  \upperAallalens$ with the ground-based experiments. 
Next we introduce inflationary gravitational waves, as parameterized by the tensor-to-scalar ratio r. 
The 95\% CL upper limit on \apmf{} drops again.  
This is easily understood as a splitting of the large-scale B-mode power between two positive-definite terms: r and \apmf{}.

We also propagate the limits on the observed PMF power into limits on the magnetic field strength, \bpmf. 
We do this using the scaling of Eqn.~\ref{eqn:scaling}, $\apmf \propto \bpmf^4$. 
Given this scaling, the observed upper limits with a flat prior on \apmf{} would lead to an apparent ``detection" of \bpmf. 
We therefore importance sample the chains to apply a flat prior on \bpmf. 
We find a 95\% CL upper limit of $\bpmf < \upperBplk$\,nG for \planck{} alone,
\changed{similar to} the limit found by \citet{planck15-19} of $\bpmf < 2.0$\,nG with $n_B=-2.9$ held fixed as we have done. 
\changed{Adding the ground-based polarization measurements significantly reduces the 95\% CL upper limit to $\bpmf < \upperBall$\,nG. }
\changed{Due to the inclusion of the \bicepkeck{} and POLARBEAR data, these results are slightly lower than the \planck{} + \sptpol{} upper limit of $<1.2\,$nG quoted by \citet{zucca16}.}

\begin{table}[tbh]
\begin{center}
\caption{\label{tab:param_all} Parameter Constraints}
\small
\begin{tabular}{l | c c c }
Model   & \apmf &\alens&$r$\\
\hline
\lcdm{}&&&\\
~~~+ \apmf{} &  $< \upperAall$ & \mbox{--} & \mbox{--} \\
~~~+ \apmf{} + \alens{} & $< \upperAallalens$ &$ \medAlensall \pm  \sigmaAlensall$ &  \mbox{--}\\
~~~+ \apmf{} + \alens{} + $r$&  $< \upperAallalensr$&$ \medAlensRall \pm  \sigmaAlensRall$ & $<\upperRall$\\ 
\end{tabular}
\tablecomments{ 
In the minimal model, \lcdm{} + \apmf{}, current CMB power spectra data from \planck, POLARBEAR, SPTpol, and \bicepkeck{} place a 95\% CL upper limit on the PMF power of $\apmf < \upperAall$. 
This limit tightens slightly if tensor modes ($r$) are allowed, or the amount of lensing is allowed to vary. 
In the case of $r$, the tightening is because the observed power at large angular scales is now split between PMFs and inflationary gravitational waves. 
In the case of \alens, it is because the preferred value for \alens{} is modestly above the predicted value of unity. 
Recall that the the PMF template is normalized to unity for \bpmf{} = 2.5\,nG.
} \normalsize
\end{center}
\end{table}

\section{Forecasts}
\label{sec:forecasts}

The sensitivity of CMB experiments is increasing rapidly due to the continued growth in the number of detectors. 
\citet{abazajian16} thus defines four stages of CMB experiments. 
Current experiments are called stage II. 
The B-mode measurements from these stage II experiments were  used for PMF constraints in the last section and are plotted in Fig.~\ref{fig:pmf-experiments}. 
Stage III experiments such as SPT-3G, the Simons Array, or AdvACTPol \citep{benson14,suzuki16,henderson16} have approximately ten times more detectors, and generically will start collecting data around 2017 and finish three to four years later.
In this section, we forecast the expected constraints from the EE and BB power spectrum measurements  of the stage III experiments by combining forecasts for SPT-3G and the Simons Array. 
There is also a proposal to build a stage IV experiment, CMB-S4, that would increase the detector counts by another order of magnitude and hopefully begin taking data in the early 2020s. 
A pathfinder to CMB-S4, the Simons Observatory, was funded in 2016.
We examine the likely PMF constraints from the EE and BB power spectrum measurements with CMB-S4, and look at how to design CMB-S4 to maximize its potential for PMF searches.

\subsection{Methods and Experimental Parameters}

We use Fisher matrices to forecast the potential constraints on PMFs possible for each generation of experiments. 
The Fisher matrix, $\mathcal{F}_{ij}$, can be defined as:
\be
\mathcal{F}_{ij} = \frac{1}{2} \frac{\partial^2 \chi^2}{\partial p_i \partial p_j},
\ee
for model parameters \textbf{p}. 
The inverse of the Fisher matrix is the covariance matrix of the parameters for Gaussian likelihoods. 
While Fisher matrices can underestimate the true uncertainties for non-Gaussian likelihoods \citep[see e.g.,][]{wolz12}, they 
 make it easy to combine information from different experiments and estimate the final parameter constraints.  

We include two external datasets in all forecasts. 
The first dataset is the expected measurements of the  TT, TE and EE spectra from the \planck{} satellite. 
We include \planck{} TT information in the multipole range $2\le \ell \le 3000$. 
Due to the importance of galactic foreground removal at large scales in polarization, we only use TE and EE information starting from $\ell = 30$. 
A prior on the optical depth of 0.005 is added to account for the optical depth constraint expected from the missing multipoles. 
Second, we include a 1\% external measurement of the Hubble constant, such as is expected from the Taipan experiment \citep{kuehn14}.

To calculate the Fisher matrix for a CMB experiment, we need the bandpower covariance matrix and partial derivatives of the CMB spectra. 
The partial derivatives are calculated numerically from CAMB spectra  \citep{lewis00}. 
We calculate the bandpower covariance matrix using the Knox formula \citep{knox97} for the survey parameters in Table~\ref{tab:experiments}. 
The sample variance contribution is calculated assuming the \planck{} best-fit \lcdm{} cosmology with a small gravitational wave contribution  ($r=0.01$) added. 
This is the black line in Fig.~\ref{fig:pmf-experiments}. 
We use a bandpower bin size of $\Delta\ell=50$. 
We  assume a 5\% uncertainty on the beam FWHM and a 1\% power calibration uncertainty.
In \S\ref{subsec:beamcal}, we test relaxing or tightening the beam and calibration uncertainty and find the beam and calibration uncertainties have a negligible impact on the PMF constraints.

One concern with combining Fisher matrices from different experiments is double-counting modes due to overlapping sky coverage. 
We avoid this problem in different ways in each of the CMB power spectra. 
For the TT and TE spectra, we only use the \planck{} measurements. 
We do not expect the future experiments to substantially improve upon \planck{} in these spectra which are already cosmic variance limited out to fairly high multipoles. 
We do the opposite for the EE and BB spectra. 
We do not include \planck{} BB information in any forecast, and we throw away \planck{} EE data in the overlap region by appropriately scaling the \planck{} EE uncertainties. 
We ignore overlaps between the stage-III experiments  on the basis that the sky overlap between a Chilean experiment like the Simons Array and South Pole experiment like SPT-3G will be small. 
The overlap issue is also the rationale behind using only one of the Simons Array and AdvACTpol as the two experiments are likely to have substantial sky overlap. 
The overlap in sky coverage will almost certainly be substantial between stage III and IV experiments, thus we do not include the stage III experiments, except as a prior on the polarized Poisson power, in the CMB-S4 constraints. 
This prior is only relevant in practice when the assumed CMB-S4 configuration has a larger beam size than SPT-3G. 
With these measures in place, this analysis should not double-count any modes.

\begin{table*}[tbh]
\begin{center}
\caption{\label{tab:experiments} Assumed survey parameters}
\small
\begin{tabular}{l || c c c c c }
Experiment & Sky coverage & Polarized Noise level  & 1/$f$ knee & Beam FWHM \\
& &($\mu$K-arcmin)&&(arcmin.)\\
\hline
\tiny \\ \small
CMB Stage III & & & & \\
~~~~~SPT-3G & 6\% & 3.0 & 200 & 1.2 \\
~~~~~Simons Array & 36\% & 9.5 & 200 & 3.5 \\ 
\tiny \\ \small
CMB Stage IV & 55\% & 1.3 & 100 & 4.0 \\
\end{tabular}
\tablecomments{ 
Key numbers about the planned stage III and IV experiments. 
The sky coverage percentages are after galactic cuts. 
Unless otherwise noted,  the Fisher matrix forecasts in this work use these numbers. 
All forecasts also allow for beam and calibration uncertainties as noted in the text. 
} \normalsize
\end{center}
\end{table*}

We consider the constraints on PMFs in two cosmologies. 
Our fiducial cosmology is a \changed{12}-parameter model that extends \lcdm{} with  four  commonly considered extensions as well as PMFs:  \lcdm{} +  $r$ + \nrun{} +  \neff{} + \mnu{} + \apmf{} \changed{ + $\beta$}. 
Here, $r$ is the tensor-to-scalar ratio, \nrun{} is running of the scalar index, \neff{} is the effective number of relativistic species, and \mnu{} is the sum of the neutrino species. 
This \changed{12}-parameter model is our default cosmological model when forecasting future PMF constraints as these extensions are all well-motivated theoretically. 
We have examined the degree to which \apmf{} is degenerate with the 10 parameters unrelated to PMFs -- the only strong degeneracy is in the $n_B\,=\,-2.9$ case and is with the tensor-to-scalar ratio, $r$. 
The reason for this degeneracy is illustrated in Fig.~\ref{fig:pmf-bb}, which shows that the tensor mode power due to inflationary gravitational waves and a PMF with $n_B\,=\,-2.9$ is nearly identical on large angular scales. 
To test the degree to which parameter degeneracies limit the inferred constraints,  we also quote constraints from a `minimal' \changed{8}-parameter model in which the PMF power is the only extension to \lcdm{}:  \lcdm{} + \apmf{} \changed{ + $\beta$}.
In both cases, we always marginalize over unknown Poisson EE and BB  terms due to polarized extragalactic sources. 
We assume the galactic foregrounds are removed by a judicious combination of data from multiple frequencies.

In this work, we restrict ourselves to power spectrum (i.e.~2-point estimators) searches for PMFs. 
Currently the power spectrum limits from \planck{} or \pb{} are better than the 4-point upper limits. 
Although outside the scope of this work, it would be interesting to extend this analysis to 4-point estimators in the future. 
First, the 4-point limits should improve faster as the noise level falls. 
Second, in the case of a detection, the 4-point estimators would almost certainly come into play to learn more about the vector modes of the PMFs. 
The detection (or non-detection) of the vector PMF signal in the 4-point estimators could then be used to argue for whether it is more likely that any observed tensor power is due to  inflationary gravitational waves or  PMFs. 
Finally, a lensing estimator might be used to `de-lense' the B-mode power spectrum, thereby suppressing the lensed B-mode signal and allowing better limits on PMFs. 
Such de-lensing techniques have long been proposed for inflationary gravitational wave searches \citep[e.g.][]{knox02,kesden02,seljak04a,simard15} and more recently been demonstrated on real \planck{} data \citep{larsen16}.

\subsection{Stage III forecasts}

The experiments that will begin taking data in 2017 will dramatically improve our knowledge of PMFs. 
In the minimal cosmological model, the 1-$\sigma$ forecast for \planck{}+\ho{} is $\sigma(\apmf)=\fisherAplk$ for a best-motivated, nearly-scale invariant PMF spectrum (i.e.~$n_B=-2.9$). 
\changed{We have set a Gaussian prior on $\beta$ with $\sigma_\beta$ chosen such that the Fisher forecast matches the actual upper limit from the \planck{} data.  }
Adding  EE and BB bandpowers from two stage-III experiments, \sptnew{} and \simons{}, reduces the uncertainty by \changed{more than} an order of magnitude in the \changed{8}-parameter model to $\sigma(\apmf)=\fisherASThree$. 
The relative improvement is larger in the more realistic \changed{12}-parameter model, as parameter degeneracies substantially weaken (by a factor of five) the \planck+\ho{} constraints on \apmf{}, while weakening the stage III constraint by a  more modest 40\%. 
Thus within the \changed{12}-parameter model,  the addition of stage III CMB experiments improves the \apmf{} uncertainty by a factor of $\sim$\,35 to  $\sigma(\apmf)=\fisherAlooseSThree$. 
The parameter degeneracies largely disappear for steeper PMF spectra (i.e. $n_B=-1$ or 2) as the PMF B-mode spectra then peaks at very small scales and this small-scale power can not be mimicked by any of the \changed{other} parameters. 
The improvement from adding the SPT-3G and Simons Array experiments to \planck{} remains impressive for these values of $n_B$.
For the \changed{12}-parameter model and $n_B=-1$, adding SPT-3G and Simons Array to \planck{} leads to a 17-fold reduction in $\sigma(\apmf)$ from \fisherAlooseNbOneplk{} to \fisherAlooseNbOneSThree{}.
For $n_B=2$, there is a 20-fold reduction from $\sigma(\apmf) = \fisherAlooseNbTwoplk$ to  \fisherAlooseNbTwoSThree. 
We can expect substantially tighter constraints on PMFs by the end of the decade. 

\subsection{Stage IV forecasts}

The primary motivations behind the proposed CMB-S4 experiment are to search for inflationary gravitational waves and to measure the neutrino masses.
However, CMB-S4 would also enable extremely sensitive searches for PMFs. 
We begin by considering our fiducial CMB-S4 configuration, as laid out in Table~\ref{tab:experiments}. 
For this fiducial configuration and the \changed{12}-parameter cosmological model, we find a 3-fold reduction over the stage III experiments for all three PMF template shapes. 
The expected uncertainties for CMB-S4 are $\sigma(\apmf^{n_B=-2.9}) =   \fisherAlooseSFour$, $\sigma(\apmf^{n_B=-1}) =   \fisherAlooseNbOneSFour$, and $\sigma(\apmf^{n_B=2}) =   \fisherAlooseNbTwoSFour$. 
These represent more than a 50-fold improvement on current limits.

Given that CMB-S4 is being designed, it is worth considering how design decisions would affect the final PMF constraints. 
We look at five aspects of the experiment: the instrumental sensitivity (as reflected by the final noise level in the maps), the telescope size or beam FWHM,  the low-$\ell$ noise performance, the amount of sky surveyed,  and how well the beam shape and calibration must be known. 
Of these five aspects, we find the first two to be very important to PMF searches and the second two to be somewhat important. 
As might be expected, in some cases the optimal design depends on the shape of the PMF spectrum ($n_B$) as this sets  the relative power between large and small angular scales.  
However, as was discussed in \S\ref{sec:template},  there are theoretical reasons to expect the PMF spectrum to be close to a scale-invariant spectra. 
While for completeness we present results for three values of $n_B$, we recommend favoring the optima for $n_B=-2.9$. 

\subsubsection{Instrumental sensitivity (map noise levels)}

\begin{figure}[htb]\centering
\includegraphics[width=0.45\textwidth,clip,trim={2.cm 12.5cm 11cm 7.5cm}]{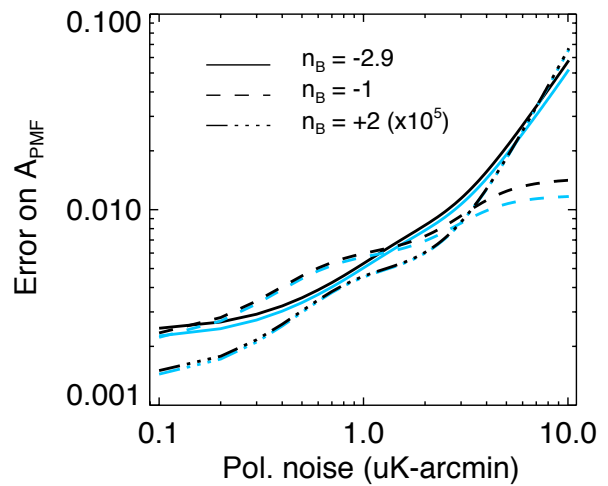}
  \caption[]{ \label{fig:sensitivity}
  How the PMF uncertainty  scales with the polarized map noise. 
  Forecasts are shown as a function of the map noise in \ukarcmin{} for each of the three PMF spectral indices: $n_B = -2.9$ (solid), -1 (dashed), and +2 (dash-dot). 
  In the case of $n_B=2$, the uncertainties have been multiplied by a factor of $10^5$ to allow them to be plotted on the same scale. 
  The black lines are for the fiducial \changed{12}-parameter model, while the light blue lines are for the minimal \lcdm{}+\apmf{} model. 
  The model only matters in the nearly-scale invariant PMF case, where the constraints degrade in the \changed{12}-parameter model due a degeneracy with the tensor-to-scalar ratio. 
  The PMF constraints improve rapidly with lower noise levels in all six cases up to $\sim$\,1\,\ukarcmin, which is close to the  fiducial CMB-S4 noise levels. 
    }
\end{figure}

Reducing the noise in CMB maps will monotonically improve power spectrum measurements. 
The goal of this section is to quantify the magnitude of that improvement: has the information gain saturated, or will further reductions in noise substantially improve PMF searches. 
Figure~\ref{fig:sensitivity} shows the uncertainty on \apmf{} as a function of the map noise level for the three PMF templates in either the \changed{12} or \changed{8}-parameter cosmological models. 
For this figure, we have fixed the rest of the experimental configuration (i.e.~everything except the polarized noise levels) to the values of the CMB-S4 row in Table~\ref{tab:experiments}. 
Clearly, continuing to improve the sensitivity of CMB experiments to CMB-S4 and beyond will be a major boon to PMF searches. 
In the general, \changed{12}-parameter cosmological model, the PMF uncertainty does not plateau in any of the three PMF templates considered until the map noise is at or below $0.3\,\ukarcmin$ (a factor of four lower then the fiducial CMB-S4 configuration).

\subsubsection{Size of Telescope }

\begin{figure}[htb]\centering
\includegraphics[width=0.45\textwidth,clip,trim={2.cm 12.5cm 11cm 7.5cm}]{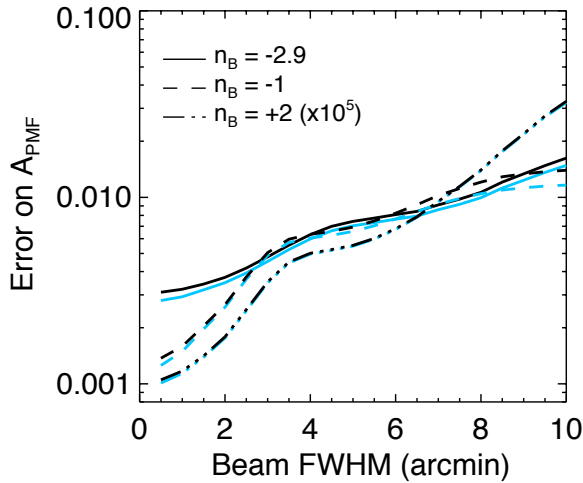}
  \caption[]{ \label{fig:beam}
 Larger telescopes (and thus smaller beams) improve limits on the PMF power. 
 Forecasts are shown as a function of the beam FWHM in arcminutes for each of the three PMF scalar indices: $n_B = -2.9$ (solid), -1 (dashed), and +2 (dash-dot). 
   In the $n_B=2$ case, the uncertainties have been multiplied by a factor of $10^5$ for plotting purposes. 
  The black lines are for the fiducial \changed{12}-parameter model, while the light blue lines are for the minimal \lcdm{}+\apmf{} model. 
  The cosmological model only matters in the nearly-scale invariant PMF case ($n_B=-2.9$), where the information on large angular scales dominates and the telescope size becomes less important. 
  The plateau near 4$^\prime$ for $n_B=-1, +2$ is where the gain from the $\ell\sim3000$ peak is saturated, but the resolution isn't yet adequate to pick up the next peak at $\ell > 8000$ (see  Fig.~\ref{fig:pmf-nb}).  
    }
\end{figure}

Larger telescopes can recover more modes on the sky, and should therefore always improve the PMF constraints. 
As shown in Fig.~\ref{fig:beam}, we find the gains due to resolution at fixed mapping speed to be substantial for all three power law indices considered. 
We see a local plateau between FWHMs of 4$^\prime$ to 6$^\prime$; improving the angular resolution in this range does little to aid PMF searches. 
We can understand this plateau by looking on the predicted PMF power in Fig.~\ref{fig:pmf-nb}. 
All three templates have two peaks across this range of angular multiples, and we would expect to see such a plateau when, for instance, the resolution is adequate to resolve the first peak, but not yet sufficient to resolve the second peak. 
The observed plateau positions are consistent with this hypothesis. 
 Overall, a beam size of FWHM=1$^\prime$ instead of 10$^\prime$ reduces the expected upper limit five-fold for $n_B=-2.9$ and 30- to 35-fold for the steeper PMF indices which are more heavily weighted towards small angular scales. 
   The model only matters in the nearly-scale invariant PMF case ($n_B=-2.9$), where the information on large angular scales dominates and the telescope size is relatively unimportant.

However, larger telescopes are also more expensive to build which means for a fixed experimental budget, they would necessitate less ambitious focal planes and lower instantaneous mapping speeds. 
We very crudely approximated a cost-neutral setup with a noise level of 1.3 \ukarcmin{} for a FWHM of 4$^\prime$; a noise level of 2.8 \ukarcmin{} for a FWHM of 2$^\prime$; and a noise level of 4.0 \ukarcmin{} for a FWHM of 1$^\prime$. 
We find that 
improving the mapping speed rather than telescope size leads to somewhat better potential limits for these experimental configurations, by a factor of 1.6, 1.1 or 1.1 for $n_B = -2.9, -1, +2$ respectively. 
However the results are close enough that the crudeness of the cost-neutral estimates used here is a worry, and the preference might flip for more accurately costed setups. 
In short,  the tradeoff between larger telescopes or more detectors is largely a wash, although one might lean towards adding more detectors. 

\subsubsection{Survey Area}

\begin{figure}[htb]\centering
\includegraphics[width=0.45\textwidth,clip,trim={2.cm 12.5cm 11cm 7.5cm}]{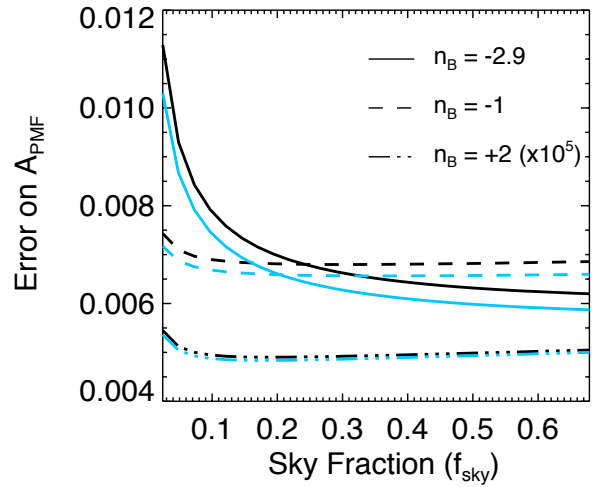}
  \caption[Area dependence]{
  The preferred survey area for PMF searches depends on the PMF spectral index. 
    In the best-motivated, nearly scale-invariant $n_B = -2.9$ case (solid line), we find PMF searches should observe more than $\sim$\,15\%. 
 In contrast, for the cases where $n_B = -1, +2$ (dashed and dot-dashed lines), we find that observing  $\sim$\,10\% of the sky produces the best PMF constraints. 
   Note that in the case of $n_B=2$, the uncertainties have been multiplied by a factor of $10^5$ for plotting purposes. 
   There is essentially no difference to the preferred areas between the \changed{12}-parameter cosmological model (black) or the restricted \lcdm{}+\changed{PMF 8}-parameter model (light blue). 
     Fortunately, the minima for all three templates are relatively broad, and a survey covering 15+\% of the sky will do well for all values of the PMF spectral index. 
    \label{fig:area}
  }
\end{figure}

A third question is how much sky to observe with CMB-S4. 
Unsurprisingly, we find opposing preferences for the nearly scale-invariant $n_B=-2.9$ (which peaks at large angular scales) and the bluer $n_B=-1$ or 2 spectra which peak at smaller scales. 
As shown in  Fig.~\ref{fig:area}, observing at least 15\% of the sky is very important to a search for nearly scale-invariant PMFs, and it is best  to cover as much sky as possible. 
A caveat to this analysis is that it is likely easier to remove galactic foregrounds to a specified level on targeted, `clean' patches as opposed to a substantial fraction of the sky, and we would expect galactic foregrounds to be important for the large angular scales. 
With that said, the PMF constraints improve by a factor of 1.4-1.5 (depending on the cosmological model) going from 10\% to 25\% of the full sky, and are nearly flat (a 6\% improvement) from 25\% to 70\% of the sky (the widest area likely to be possible after galactic cuts).
Conversely for $n_B=-1$ or +2, the optimal survey area is \changed{around 15}\% of the full sky. 
Note however that the minimum is extremely broad with only a 10\% worsening of the expected uncertainty as the survey area is increased from 10\% to 70\% of the  full sky. 
A survey \changed{ covering at least} 25\% of the sky would perform well for all three considered PMF models. 

\subsubsection{Noise performance at large angular scales}

\begin{figure}[htb]\centering
\includegraphics[width=0.45\textwidth,clip,trim={2.cm 12.5cm 11cm 7.5cm}]{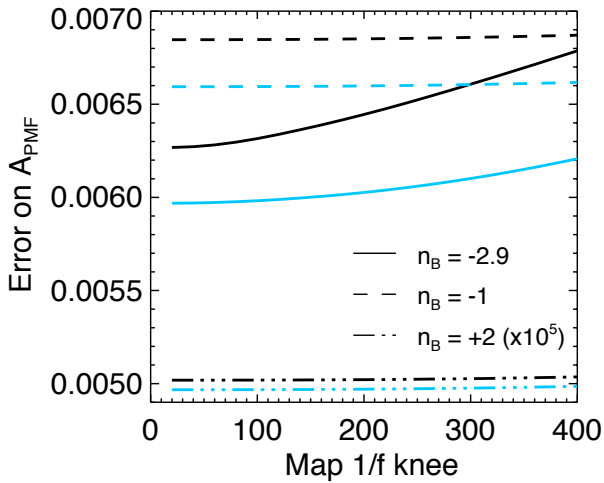}
  \caption[Map knee dependence]{
  Recovering large angular scales is only important for a nearly scale-invariant PMF ($n_B = -2.9$) in the restricted \lcdm{}+\changed{PMF 8}-parameter model (solid, light blue line). 
The achieved map 1/f knee frequency makes no difference to CMB-S4 PMF searches in the \changed{12}-parameter model (black lines). 
  For $n_B=-2.9$, this transition occurs because the \changed{12}-parameter model introduces a degeneracy with r on the large angular scales impacted by the map 1/f knee. 
  The 1/f knee frequency also makes no difference for the bluer PMF templates  (dashed and dash-dotted black lines) in either the restricted (light blue) or full \changed{12}-parameter (black) model space. 
     Note that in the case of $n_B=2$, the forecasts have been multiplied by a factor of $10^5$ for plotting purposes. 
    \label{fig:knee}
  }
\end{figure}

Next we turn to the recovery of large angular scales on the sky, and the required noise performance at these low frequencies. 
For this, we look at the impact of shifting the 1/f knee of the map-space noise. 
Specifically, we are multiplying the noise power, $N_\ell$, which is a constant for white noise,  by a function of angular multipole:
\be \label{eqn:knee}
f(\ell) = 1 + \left(\frac{\ell_{\rm knee}}{\ell}\right)^{8/3}.
\ee 
The exponent, 8/3, was selected based on a Kolmogorov spectrum of turbulence within a thin plane \citep{lay00}. 
Note that this knob  serves as a placeholder for several effects, including a signal-to-noise hit due to galactic foregrounds or the methods used to clean these foregrounds, atmospheric noise, or actual instrumental low frequency noise. 

We generally find the forecasts to be insensitive to the 1/f knee or galactic foreground removal at the default beam size of 4$^\prime$. 
It appears that even in the nearly scale-invariant case, the PMF constraint is coming primarily from smaller angular scales due to the degeneracy \changed{at large angular scales with both the tensor-to-scalar ratio and the PMF timing parameter $\beta$}. 
For the bluer PMF spectra ($n_B=-1$ or 2), the forecasts show no dependence on the knee frequency. 
For the fiducial \changed{12}-parameter model, our ability to separate a PMF from other physics is coming from sub-degree angular scales. 
The low frequency noise performance of CMB-S4 is not crucial to PMF searches.

\subsubsection{Beam and calibration uncertainties}
\label{subsec:beamcal}

Finally, we consider if searches for PMFs introduce new requirements on the accuracy to which the beam or  calibration of an experiment must be known. 
We find they do not. 
We parameterize the beam uncertainty as a fractional uncertainty on the FWHM of a Gaussian beam, and calibration uncertainty as an overall power uncertainty. 
We find negligible, sub-percent shifts in the forecasted PMF uncertainties for calibration uncertainties from 0.2 to 5\% and beam FWHM uncertainties from 2 to 12.5\%. 
The PMF constraints  for all three power law indices considered are insensitive to the beam and calibration uncertainties. 

\section{Conclusions}
\label{sec:conclusions}

In this work, we have improved the current upper limits on the strength of   primordial magnetic fields by including more CMB B-mode polarization data. 
By adding \bicepkeck{}, \pb, and \sptpol{} to \planck{} we have found that the 95\% CL upper limit on the PMF power falls nearly two-fold from $\apmf < \upperAplk$ to $\apmf < \upperAall$. 
\changed{The two major contributors to this improvement are the low-$\ell$ B-mode data from \bicepkeck{} and the high-$\ell$ B-mode data from \sptpol. }

We have also shown that the next generation of experiments should dramatically reduce these limits, with the potential to detect PMFs for the first time. 
The so-called stage III experiments, which will begin taking data in 2017, can be expected to set upper limits at the level of $\apmf < 0.02$ even after marginalizing over a six-parameter extension to \lcdm{} and foregrounds. 

The potential for detection increases even further with planned experiments like the Simons Observatory or CMB-S4. 
We have shown that an ideal version of CMB-S4 might decrease the 95\% CL upper limits eight-fold, to $\apmf < 0.0025$, and \changed{more realistic} versions can still set upper limits on order of $\apmf < 0.006$ in a \changed{12}-parameter cosmological model with  \lcdm{} +  $r$ + \nrun{} +  \neff{} + \mnu{} + \apmf{} \changed{+$\beta$}. 
This represents a three-fold improvement over the forecasts for the stage III experiments and a more than 100-fold improvement over the current limits for this cosmological model. 
We have considered how the design of future experiments will impact the resulting PMF limits, looking at the experimental sensitivity, angular resolution, survey area, low-frequency noise performance, and beam or calibration uncertainties. 
We have found an experiment's sensitivity followed by telescope size to be the most important factors in predicting the PMF limits. 
CMB-S4 will be a very exciting probe of PMFs and other sources of cosmic birefringence. 

\acknowledgments

We thank the referee as well as Srinivasan Raghunathan and Federico Bianchini for valuable feedback on the manuscript. 
CR is the recipient of an Australian Research Council Future Fellowship (FT150100074), and also acknowledges support from the University of Melbourne. 
CF acknowledges support from NASA grants NASA NNX16AJ69G and NASA NNX16AF39G. 
This research used resources of the National Energy Research Scientific Computing Center, which is supported by the Office of Science of the U.S. Department of Energy under Contract No. DE-AC02-05CH11231. 
We acknowledge the use of the Legacy Archive for Microwave Background Data Analysis (LAMBDA). Support for LAMBDA is provided by the NASA Office of Space Science.

\bibliography{pmf}

\end{document}